\documentclass[aps,prl,reprint, twocolumn,superscriptaddress,fleqn]{revtex4-1}
\usepackage{soul}
\usepackage{graphicx}
\usepackage{amsmath, amsfonts, amssymb, amsbsy}
\usepackage{MnSymbol}
\usepackage[usenames,dvipsnames]{xcolor}
\usepackage{marvosym}
\usepackage{sidecap}
\usepackage{ifsym}
\usepackage{color}

\newcommand{\beq} {\begin{equation}}
\newcommand{\eeq} {\end{equation}}

\newcommand{\cb} {{c\parallel B_0}}
\newcommand{\cp} {{c\bot B_0}}
\newcommand{\tc} {{T_{\rm c}}}
\newcommand{\lsco} {{La$_{2-x}$Sr$_x$CuO$_4$}} 
 
\newcommand{\ybco} {{YBa$_{2}$Cu$_3$O$_{6+\delta}$}}

\newcommand{\chic} {{\chi_{\rm 12}}}

\newcommand{\aus}[1]{}
\begin{document}


\title{Electronic spin susceptibilities and superconductivity in HgBa$_{2}$CuO$_{4+\delta}$ from nuclear magnetic resonance}


\author{Damian Rybicki}
\affiliation{University of Leipzig, Faculty of Physics and Earth Sciences,  Linnestr. 5, 04103 Leipzig, Germany}
\affiliation{AGH University of Science and Technology, Faculty of Physics and Applied Computer Science,
Department of Solid State Physics, al. A. Mickiewicza 30, 30-059 Krakow, Poland}
\thanks{Corresponding author}
\email{ryba@agh.edu.pl}
\author{Jonas Kohlrautz}
\affiliation{University of Leipzig, Faculty of Physics and Earth Sciences,  Linnestr. 5, 04103 Leipzig, Germany}
\author{J\"urgen Haase}
\affiliation{University of Leipzig, Faculty of Physics and Earth Sciences,  Linnestr. 5, 04103 Leipzig, Germany}
\author{Martin Greven}
\affiliation{School of Physics and Astronomy, University of Minnesota, Minneapolis, Minnesota 55455, USA}
\author{Xudong Zhao}
\affiliation{School of Physics and Astronomy, University of Minnesota, Minneapolis, Minnesota 55455, USA}
\affiliation{College of Chemistry, Jilin University, Changchun 130012, China}
\author{Mun K. Chan}
\altaffiliation[Present address: ]{National High Magnetic Field Laboratory, Los Alamos National Laboratory, Los Alamos, New Mexico 87545, USA}
\affiliation{School of Physics and Astronomy, University of Minnesota, Minneapolis, Minnesota 55455, USA}
\author{Chelsey J. Dorow}
\altaffiliation[Present address: ]{Department of Physics, University of California, San Diego, La Jolla, California 92093, USA}
\affiliation{School of Physics and Astronomy, University of Minnesota, Minneapolis, Minnesota 55455, USA}
\author{Michael J. Veit}
\altaffiliation[Present address: ]{Department of Applied Physics, Stanford University, Stanford, California 94305, USA}
\affiliation{School of Physics and Astronomy, University of Minnesota, Minneapolis, Minnesota 55455, USA}
\date{\today}

\begin{abstract}
Nuclear magnetic resonance (NMR) experiments on single crystals of HgBa$_{2}$CuO$_{4+\delta}$ are presented that identify two distinct temperature-dependent spin susceptibilities: one is due to a spin  component that is temperature-dependent above the critical temperature for superconductivity ($\tc$) and reflects pseudogap behavior; the other is Fermi-liquid-like in that it is temperature independent above $\tc$ and vanishes rapidly below $\tc$. In addition, we demonstrate the existence of a third, hitherto undetected spin susceptibility: it is temperature independent at higher temperatures, vanishes at lower temperatures (below $T_0\neq T_{\rm c}$), and changes sign near optimal doping. This susceptibility either arises from the coupling between the two spin components, or it could be given by a distinct third spin component.
\end{abstract}




\maketitle

\begin{figure}
   \centering
   \includegraphics[width=0.4\textwidth]{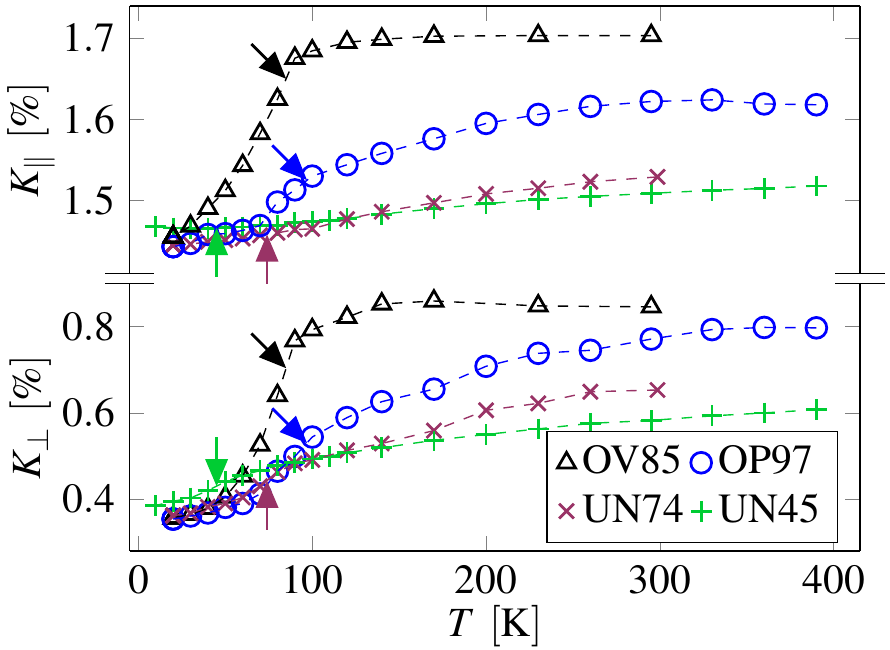}
   \caption{(Color online) Total magnetic ${^{63}}$Cu shifts $K_\eta$ as a function of temperature. Upper panel: $B_0$ parallel to the crystal $c$-axis ($K_\parallel$); lower panel: $B_0$ in the CuO$_2$ plane ($K_\bot$). For $K_\bot$, the contribution from the quadrupole interaction was removed. Dashed lines are guides to the eye. Arrows indicate $\tc$ values. Errors are smaller than the data point size.}
\label{fig:sT1}
\end{figure}

The high-temperature superconducting cuprates have long been known to exhibit complex electronic properties, yet there exists increasing evidence for Fermi-liquid behavior throughout much of the phase diagram. NMR is a powerful probe of local electronic degrees of freedom, and the complex properties of the cuprates manifest themselves in a unique dependence of NMR shifts on temperature and doping. The data in Fig.~\ref{fig:sT1} are a good example: at high doping levels \textit{and} high temperatures, the shifts are rather independent of temperature, and they rapidly decrease below $\tc$ (reminiscent of a Fermi liquid). As the doping level is lowered, the pseudogap makes the shifts temperature-dependent even above $\tc$, whereas the sudden decrease below $\tc$ disappears.

A long-standing fundamental question has been whether a single electronic fluid's temperature-dependent electronic spin polarization, $S(T)= \chi(T) B_0$, in a magnetic field, $B_0$, can explain these shifts. From the analyses of YBa$_{2}$Cu$_3$O$_{6.63}$ and YBa$_{2}$Cu$_4$O$_{8}$ shifts measured at planar copper and oxygen above and below $\tc$ it was concluded that this is the case \cite{Takigawa1989, Bankay1994}. 
Henceforth, NMR data have been interpreted largely in terms of a single electronic spin component. 
In contrast, early uniform susceptibility measurements above $\tc$ concluded on the presence of two spin components: a temperature-dependent component ("pseudogap-like"), and a temperature-independent component (Fermi-liquid-like) \cite{Johnston1989, Nakano1994}.

NMR shift experiments are rather reliable since $\chi(T)$ must cause proportional spin shifts for \textit{all} nuclear resonances. For a given orientation ($\eta$) of a crystal with respect to $B_0$, we expect a spin shift $K_{S\eta}(T) = {q_\eta} \cdot \chi(T)$, where the anisotropy arises only from the effective hyperfine coefficients of each nucleus (${q_\eta}$) since $\chi(T)$ is believed to be isotropic.

A few years ago, it was shown that the spin shifts at Cu and O in La$_{1.85}$Sr$_{0.15}$CuO$_4$ cannot be explained with a single spin component's $\chi(T)$, but rather require two spin components with distinct temperature dependencies \cite{Haase2009}. One of the components, $S_1(T)$, causes the pseudogap response, and it dominates the planar O shift. The second component, $S_2(T)$, is temperature independent above $\tc$ and  rapidly vanishes below it, reminiscent of Fermi liquid behavior. The second component dominates the planar Cu and apical O shifts. 

Two spin components, $S_{1}$ and $S_{2}$, affect a nucleus through $q_{1\eta}$ and $q_{2\eta}$, respectively, so that its spin shift is 
\beq
K_{S\eta}(x,T) = {{q}_{1\eta}} \cdot \chi_{1}(x,T)+{{ q}_{2\eta}} \cdot \chi_{2}(x,T).
\label{eq:shift10}
\eeq
We note that, if $S_{1}$ and $S_{2}$ are coupled, $\chi_{1}$ and $\chi_{2}$ must be the sum of two terms each, i.e., $\chi_1=\chi_{11}+\chic$  and $\chi_2=\chic+\chi_{22}$, where $\chic$ is the coupling susceptibility that describes how $S_1$ responds to a magnetic field acting on $S_2$ \cite{Curro2004, Haase2009, Haase2012}. If we investigate just one nucleus for different orientations of $B_0$, Eq.~\eqref{eq:shift10} also holds if the anisotropy of $q_{1\eta}$ is different from that of $q_{2\eta}$.

Motivated by experiments on La$_{1.85}$Sr$_{0.15}$CuO$_4$, we subsequently investigated another single-layer system, HgBa$_{2}$CuO$_{4+\delta}$. With ${^{63}}$Cu and ${^{199}}$Hg NMR on underdoped ($\tc$=74 K, UN74) and  optimally doped ($\tc$=97 K, OP97) single crystals, the failure of a single component approach became apparent as well \cite{Haase2012}. However, the doping dependence of the temperature independent component remained unclear \cite{Haase2012}. The reason for this will be uncovered here. We confirm shift components due to $S_1$ and $S_2$, but we also discover a new shift component that is temperature independent at high temperatures and vanishes at low temperatures. However, it differs from the Fermi-liquid-like component in that it changes sign as a function of doping (it is nearly zero for optimal doping), furthermore, the characteristic temperature ($T_0$) at which it suddenly begins to disappear, depends only weakly on doping and can be larger than $\tc$ for underdoped, and lower than $\tc$ for overdoped samples. Since $T_0$ is similar to $\tc$ for UN74 this component was not identified earlier \cite{Haase2012}. We argue below that this new component is likely a generic property of all cuprates.

Two new HgBa$_{2}$CuO$_{4+\delta}$ single crystals with $\tc$=45 K (UN45) and 85 K (UN85) were prepared following the method described previously \cite{zhao2006,Barisic2008}. The experimental details of exciting, recording and referencing the $^{63}$Cu NMR signals are identical to those in Ref.~\cite{Haase2012}. In Ref.~\cite{Haase2012} it was also shown that the diamagnetic response due to the mixed state below $\tc$ can be neglected for $^{63}$Cu shifts, making them very reliable also below $\tc$.

In Fig.~\ref{fig:sT1}, we show the measured $^{63}$Cu shifts, $K_{\parallel}(T)$ and $K_{\bot}(T)$, for all HgBa$_{2}$CuO$_{4+\delta}$ single crystals studied (including those from Ref. \cite{Haase2012}). We display the total experimentally measured magnetic shift, $K_\eta(T)=K_{L\eta}+K_{S\eta}(T)$, which is the sum of a temperature and doping {\it independent} orbital part ($K_{L\eta}$) \cite{Renold2003} and the temperature and doping {\it dependent} spin part ($K_{S\eta}$).

\begin{figure}
   \centering
   \includegraphics[width=0.4\textwidth]{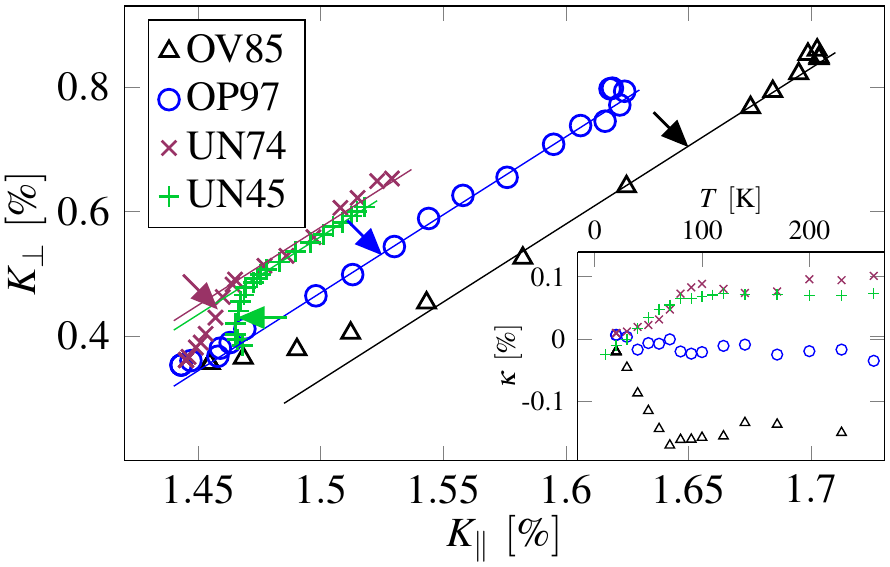}
   \caption{(Color online) $K_{\perp} (T)$ versus $K_{\parallel} (T)$ with temperature as an implicit parameter. Arrows indicate $\tc$ values. The straight lines have the slope 2.5 obtained from the fit to the data down to $T_0$. Inset shows $K_{S,\perp} (T) - 2.5 K_{S,\parallel} (T)$ as a function of temperature.}
\label{fig:ss01}
\end{figure}

In Fig.~\ref{fig:ss01}, we show the same data, but plotted as $K_\bot(T)$ versus $K_\parallel(T)$. At larger temperatures (large shift values) parallel lines appear that begin to approach a common low-temperature point below a characteristic temperatures $T_0 \neq \tc$ (cf.~Tab.~\ref{tableTzero}). This implies the presence of a shift component that is temperature-independent at high temperatures, but disappears below $T_0$. With just the data for UN74 and OP97 it was erroneously concluded  \cite{Haase2012} that this offset between the parallel lines is due to the Fermi-liquid-like component. In  order to analyze the data in Fig.~\ref{fig:ss01} we write
\beq
{K}_{S\bot}(T) =\frac{1}{c_0}{K}_{S\parallel}(T)+\kappa(x,T),
\label{eq:shiftex}
\eeq
where  $\kappa(x,T)$ describes the temperature dependent offset between the parallel lines, which is plotted in the inset in Fig.~\ref{fig:ss01}. We adopt the typical definition of the spin shift, $K_{S\eta}$, by choosing $K_{L\eta}$ as the remaining shift at the lowest temperatures, i.e., ${K}_{S\eta}(T)=K_{\eta}(T)-K_{L\eta}$, but the basic findings do not depend on the choice of $K_{L\eta}$ (that is why we show the total shifts in Figs.~\ref{fig:sT1}, \ref{fig:ss01}). From the slopes we determine $c_0\approx 0.40\pm0.02$.
We are certain that the new shift component is due to a spin susceptibility ($\chi_\kappa$), i.e., $ \kappa(x,T) \propto \chi_\kappa $. First, it is  natural to associate a temperature-dependent shift with electronic spin, and second,  we find evidence for $\chi_\kappa$ also in $^{199}$Hg NMR \cite{Haase2012}, as well as recent $^{17}$O NMR  \cite{Mounce2013} (see Supplement). 
\begin{table}
  \centering
\begin{tabular}{c|cccc|cc}
  &$x$&$T_0$&&&$x$&$T_0$\\
  \cline{1-3}
  \cline{5-7}
  &&&&&&\\[-3mm]
  UN45&0.06(1)&80(10)~K&\quad\quad\quad&OP97&0.16(2)&$\approx$75(10)~K\\
  UN74&0.10(1)&80(10)~K&\quad&OV85&0.19(1)&60(10)~K
\end{tabular}
\caption{Values of doping level $x$ \cite{Barisic2008} and $T_0$.}
\label{tableTzero}
\end{table}

\begin{figure}
   \centering
   \includegraphics[width=0.4\textwidth]{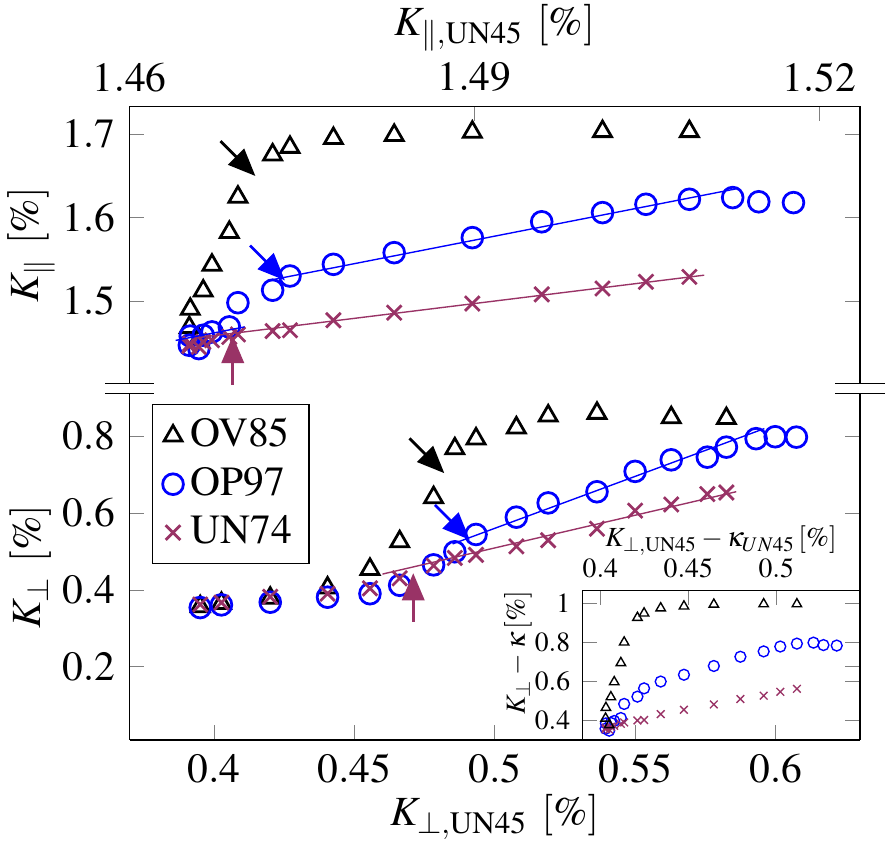}
   \caption{(Color online) $K_\parallel$ (upper panel) and $K_\bot$ (lower panel) of the UN74, OP97 and OV85 samples plotted versus shifts of the UN45 sample with temperature as an implicit parameter. Straight lines have slopes derived from doping ratios. Inset shows $K_\bot - \kappa$ of UN74, OP97 and OV85 samples versus $K_\bot - \kappa$ of the UN45 sample.}
\label{fig:ssds}
\end{figure}

We can learn more about the spin components and their susceptibilities just from the highly reliable Cu shifts. As reported earlier \cite{Rybicki2009,Haase2012}, the pseudogap shift component ($K_{S,PG}$) has a unique temperature dependence, at least up to optimal doping: $K_{S,PG}(x,T)=x\cdot \sigma(T)$, where $x$ is the average doping level of the sample and $\sigma(T)$ a universal function of temperature. Our new data support this scaling, and we explain in more detail in the Supplement that this scaling behavior is even in quantitative agreement with early susceptibility data  \cite{Johnston1989, Nakano1994}  for the pseudogap susceptibilities of other cuprates. As a consequence, if one plots the shifts measured on samples with different doping levels against each other (with temperature as an implicit parameter),  straight lines or line segments are found. This can be seen in Fig.~\ref{fig:ssds}, and indeed, the slopes of the linear segments are equal to the doping ratios.
(It is worth noting that a similar scaling was also observed for the electronic entropy of YBa$_2$Cu$_3$O$_{6+\delta}$ and Bi$_2$Sr$_2$CaCu$_2$O$_{8+\delta}$ \cite{Storey2012}.)

We now discuss Fig.~\ref{fig:ssds} in more detail.
First, we consider UN45 and UN74. 
For $\cb$, the shifts for these two samples are nearly proportional to each other (throughout the whole temperature range). This is true for $\cp$ as well after subtracting $\kappa(x,T)$ (cf. inset in Fig.~\ref{fig:ssds}). With the proportionality of the two shifts, not interrupted near either sample's $\tc$, we conclude that the shift due to $S_2$ must be negligible.

Next, we examine OP97 (for which $\kappa \approx 0$, cf.~Fig.~\ref{fig:ss01}). As concluded earlier \cite{Haase2012}, in a broad temperature range above and below $\tc$ we find the expected slope for both orientations (Fig.~\ref{fig:ssds}). 
The sudden change of $K_{\parallel, \rm OP97}$ near 97~K must then be due to $S_2$. (The decrease starts at $\tc$ with $\Delta K_{\parallel,{\rm OP97}}={ q}_{1\parallel} \Delta\chi_{{\rm OP97}}\approx 0.05\% $  and is completed at $T\approx$ 75~K. 
For $\cp$ we find $\Delta K_{\bot,{\rm OP97}}={q}_{1\perp} \Delta\chi_{{\rm OP97}}\approx 0.13\%$, in agreement with the ratio $c_0=q_{1\parallel}/{q_{1\perp}}$.) This means that the anisotropies of the hyperfine coefficients for both spin components, $S_1$ and $S_2$, are the same, so that the corresponding changes in the shifts do not show any discontinuities in Fig.~\ref{fig:ss01}. 

We now turn to OV85. Going back to Fig.~\ref{fig:sT1}, we notice that $K_\eta(T)$ is nearly constant above $\tc$, but starts to rapidly decrease  at $\tc$ (as if dominated by $S_2$). Fig.~\ref{fig:ss01} reveals that this decrease begins well above the temperature $T_0$ below which $\kappa(T)$ begins to change (i.e., when the slope in Fig.~\ref{fig:ss01} changes). Again, this says that the two shift components due to $S_1$ and $S_2$ share the same anisotropy of the hyperfine coefficients.

\begin{figure}
   \centering
   \includegraphics[width=0.4\textwidth]{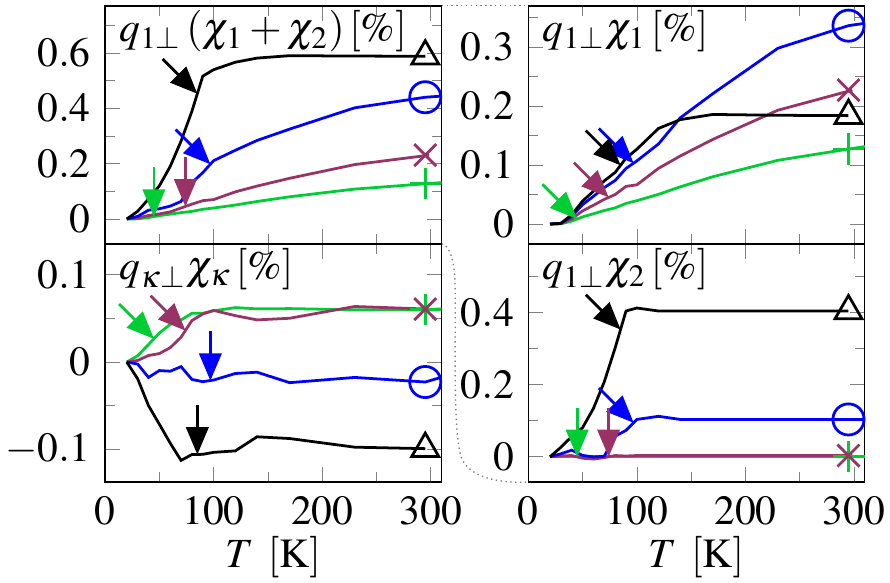}
   \caption{(Color online) Results of the numerical decomposition of the spin shifts 
for $\cp$. Left panel: into ${q_{1\bot} \left(\chi_{1}+\chi_2\right)}$ and ${q_{\kappa\bot} \chi_{\kappa}}$ according to \eqref{eq:newshift10}. Right panel: into the pseudogap component ${q_{1\bot} \chi_{1}}$ and the Fermi-liquid-like component ${q_{1\bot} \chi_{2}}$. The arrows indicate $\tc$ values, symbols are only to help identify the samples.}
\label{fig:sketch}
\end{figure}

To conclude, we have identified three spin shift components that differ in their temperature and doping dependence, and since two of them share the same anisotropy we analyze all shifts with the following simple model,
\beq
K_{S\eta}(x,T) = {{q}_{1\eta}} \left[\chi_{1}(x,T)+ \chi_{2}(x,T)\right]+q_{\kappa\eta}\chi_{\kappa}(x,T).
\label{eq:newshift10}
\eeq
In this analysis, we assume that (1) the pseudogap shift is caused by $\chi_1$ that obeys the scaling behavior discussed above; (2) for UN45 and UN74, the shifts are given by $\chi_1$ and $\chi_\kappa$ since there are no shift changes at $\tc$; and (3) that $\chi_2$ is constant above $\tc$. This leads to the results displayed in Fig.~\ref{fig:sketch} for $\cp$ (the results for $\cb$ differ only in magnitude due to anisotropy of $\rm q_{1\eta}$ and $\rm q_{\kappa\eta}$). A  detailed description of the analysis is given in the Supplement.

The left panel of Fig.~\ref{fig:sketch} shows the first step of the decomposition: we see how ($\chi_1+\chi_2$) and $\chi_\kappa$ evolve with temperature and doping. $\chi_\kappa$ changes sign near optimal doping and is almost twice larger in magnitude for OV85 than for the two underdoped samples. 
In the right panel of Fig.~\ref{fig:sketch} we extract $\chi_1(T)$ and $\chi_2(T)$ using the scaling of $\chi_1$.
At low doping, $\chi_{2}$ is negligible, but rapidly increases with doping. For the temperature range of our study, $\chi_1$ grows with increasing doping up to optimal doping. It can be identified even for OV85 at lower temperatures, but its high-temperature behavior cannot be reliably extracted. 

If $\chi_1$ is the susceptibility of $S_1$ and $\chi_2$ that of $S_2$, $\chi_\kappa$ could be due to the coupling between $S_1$ and $S_2$, i.e., $q_{\kappa\eta}\chi_{\kappa}(x,T)=2 q_{1\eta}\chi_{12}$. As such, the sign change of $\chi_\kappa$ with doping may indicate a change in sign of the electronic spin-spin coupling. However, since the anisotropies of $q_{1\eta}$ and $q_{\kappa \eta}$ are different, $\chi_\kappa$ would have to be anisotropic. This may not be possible in the absence of spin-orbit coupling. Alternatively, $\chi_\kappa$ could be the susceptibility of a  new spin component ($S_3$). In such a case, coupling of $S_3$ to $S_1$ and $S_2$ could possibly be leading to a complicated shift scenario that can, however, be described in a rather simple way as shown here.
We would like to point out that $\kappa$ cannot be explained by a redistribution of NMR spectral weight with temperature within the rather broad Cu resonance. This is also seen from the Hg NMR linewidth \cite{Haase2012}, since they are smaller than the changes due to $\kappa$. 

The fact that the Cu nucleus couples to $S_1$ and $S_2$ with just one anisotropic $q_{1\eta}$ is perhaps not surprising, but argues against a trivial picture of different Cu and O spins to which a Cu nucleus would couple with different angular dependencies. Perhaps, $S_1$ and $S_2$ relate to anti-nodal and nodal quasi-particles, respectively, which may be coupled to give $\chi_\kappa$ \cite{ChubukovPrivComm}. Pines and Barzykin explained the temperature and doping dependence of the uniform spin susceptibility of \lsco \; and YBa$_2$Cu$_3$O$_{6+x}$, assuming coexistence of two electronic fluids: a two-dimensional local moment spin liquid and a quasiparticle fermion liquid \cite{Barzykin2009, Pines2013}. 

While there can be no doubt about the existence of $\chi_\kappa$ in HgBa$_{2}$CuO$_{4+\delta}$, the question arises whether it is of relevance for the other cuprates as well. Certainly, such a term could be present in NMR shift data, but pass unnoticed. First, $\chi_\kappa$ is temperature independent at high temperatures and thus difficult to distinguish from orbital shifts, and from $\chi_2$; at low temperatures, the diamagnetic response in the superconducting state obscures its temperature dependence ($\tc$ is often not very well defined and thus hard to distinguish from a smeared $T_0$). In addition, broad NMR lines and signal wipe-out on the underdoped side of the phase diagram create uncertainties. It is no surprise that we discovered the new component with HgBa$_{2}$CuO$_{4+\delta}$ single crystals, as most $^{63}$Cu NMR shift studies involved systems for which  $^{63}K_{S\parallel}(T) \approx 0$ due to an accidental cancellation (e.g., \lsco, \mbox{\ybco)}.  The two-component descriptions of La$_{1.85}$Sr$_{0.15}$CuO$_4$ and YBa$_2$Cu$_4$O$_8$ \cite{Haase2009, Meissner2011} do not allow to distinguish $\chi_\kappa$ from $\chi_2$ (for La$_{1.85}$Sr$_{0.15}$CuO$_4$ a temperature $T_{\rm const}$ was introduced, but could not be reliably distinguished from $\tc$; for YBa$_2$Cu$_4$O$_8$ under high pressure only $^{17}$O data for one orientation could be recorded reliably). 

Bulk susceptibility data are only available above $\tc$, and at low temperatures the data are often obscured by a Curie-like response \cite{Johnston1989}, but nevertheless, as we show in greater detail in the Supplement,  a large body of susceptibility data on different cuprates is in agreement with our shift data \cite{Johnston1989, Nakano1994}. They clearly show, based on the scaling property of one component (the pseudogap susceptibility) that another temperature-independent component above $\tc$ must be present (the Fermi-liquid component). The latter can easily include $\chi_\kappa$. (We would like to note that in Johnston's analysis \cite{Johnston1989} $\chi_1(T=0)>0$, which should result in a non-zero NMR shift due to the pseudogap susceptibility at the lowest temperatures; in our analysis such a temperature independent shift is contained in the orbital shift). Therefore, the new component might be universal to the cuprates. 

The scenario found here reminds one of  a quantum critical point near optimal doping \cite{Chubukov2014} (where $\chi_\kappa$ changes sign): on the underdoped side we have $\chi_1$ and $\chi_\kappa$, on the overdoped side $\chi_2$ and $\chi_\kappa$. It is  not clear whether the Fermi-liquid-like behavior in the underdoped region observed in other experiments (d.c. resistivity, optical conductivity, and magnetoresistance measurements) on HgBa$_{2}$CuO$_{4+\delta}$ \cite{Barisic2013, Mirzaei2013, Chan2014} corresponds to a small Fermi-liquid-like component (invisible to NMR) or is related to $\chi_\kappa$. An important question to be addressed in future experiments is whether $\chi_\kappa$ and $\chi_2$  are perhaps connected with the normal-state charge-density-wave correlations and the quantum oscillations observed below optimal doping \cite{Barisic2013b, Doiron2013, Tabis2014}.

To conclude, based on a detailed study of the local magnetic response of HgBa$_{2}$CuO$_{4+\delta}$ single crystals we confirm that a description of the NMR shifts with a single, temperature-dependent spin component is not possible. Since this finding applies to three different classes of materials \cite{Haase2009, Meissner2011}, it must be generic for the cuprates. As reported before, one shift component is due to the pseudogap and it governs the NMR shifts at lower doping levels. The second component shows Fermi-liquid-like behavior and governs on the overdoped side of the phase diagram, where the pseudogap shift is suppressed. We discovered a new, third shift component that could not be distinguished from the Fermi-liquid-like component, earlier \cite{Haase2012}. The new component is temperature independent above a critical temperature $T_0$, which can be significantly larger than $\tc$ for underdoped or smaller than $\tc$ for overdoped crystals. Since it changes sign (near optimal doping), and it disappears below $T_0$ rather than $\tc$, it is very different from the Fermi-liquid-like component. Furthermore, the anisotropy of its hyperfine coefficient with the Cu nucleus is different from that of the pseudogap and Fermi-liquid-like components, which share the same anisotropy and thus probably the same atomic orbitals. Therefore, the new component could reflect the coupling between the pseudogap and Fermi liquid spins only if spin rotation symmetry were broken. Alternatively, it could represent a distinct third spin component. 

\begin{acknowledgments}
 
\textbf{Acknowledgement}
We are thankful to C.P. Slichter, A. Chubukov, M. Jurkutat, and Cz. Kapusta for discussions. We thank Y. Li and G. Yu for preparing the UN75 and OP97 crystals that enabled the prior work \cite{Haase2012} and present data analysis. We also acknowledge financial support by Leipzig University, the DFG within the Graduate School BuildMoNa, the European Social Fund (ESF) and the Free State of Saxony. The work at the University of Minnesota was supported by the US Department of Energy, Office of Basic Energy Sciences under Award No. DE-SC0006858.
 
\end{acknowledgments}

\bibliography{HTSC}
\bibliographystyle{apsrev4-1}

\end{document}




\title{Supplement to:\\
Electronic spin susceptibilities and superconductivity in HgBa$_{2}$CuO$_{4+\delta}$ from nuclear magnetic resonance}



\author{Damian Rybicki}
\affiliation{University of Leipzig, Faculty of Physics and Earth Sciences,  Linnestr. 5, 04103 Leipzig, Germany}
\affiliation{AGH University of Science and Technology, Faculty of Physics and Applied Computer Science,
  Department of Solid State Physics, al. A. Mickiewicza 30, 30-059 Krakow, Poland}
\thanks{Corresponding author}
\email{ryba@agh.edu.pl}
\author{Jonas Kohlrautz}
\affiliation{University of Leipzig, Faculty of Physics and Earth Sciences,  Linnestr. 5, 04103 Leipzig, Germany}
\author{J\"urgen Haase}
\affiliation{University of Leipzig, Faculty of Physics and Earth Sciences,  Linnestr. 5, 04103 Leipzig, Germany}
\author{Martin Greven}
\affiliation{School of Physics and Astronomy, University of Minnesota, Minneapolis, Minnesota 55455, USA}
\author{Xudong Zhao}
\affiliation{School of Physics and Astronomy, University of Minnesota, Minneapolis, Minnesota 55455, USA}
\affiliation{College of Chemistry, Jilin University, Changchun 130012, China}
\author{Mun K. Chan}
\altaffiliation[Present address: ]{National High Magnetic Field Laboratory, Los Alamos National Laboratory, Los Alamos, New Mexico 87545, USA}
\affiliation{School of Physics and Astronomy, University of Minnesota, Minneapolis, Minnesota 55455, USA}
\author{Chelsey J. Dorow}
\altaffiliation[Present address: ]{Department of Physics, University of California, San Diego, La Jolla, California 92093, USA}
\affiliation{School of Physics and Astronomy, University of Minnesota, Minneapolis, Minnesota 55455, USA}
\author{Michael J. Veit}
\altaffiliation[Present address: ]{Department of Applied Physics, Stanford University, Stanford, California 94305, USA}
\affiliation{School of Physics and Astronomy, University of Minnesota, Minneapolis, Minnesota 55455, USA}

\date{\today}
 \maketitle

\makeatletter 
\renewcommand{\thefigure}{S\@arabic\c@figure}
\makeatother
\makeatletter 
\renewcommand{\theequation}{S\@arabic\c@equation}
\makeatother
\section{Part 1: Detection of $\chi_\kappa(T)$}
Here we discuss why the new shift component $\kappa(x,T)$ must be due to electronic spin. We will give further evidence that proves the existence of $\chi_\kappa(x,T)$. 

\begin{figure}[h]
  \centering
  \includegraphics[width=\columnwidth]{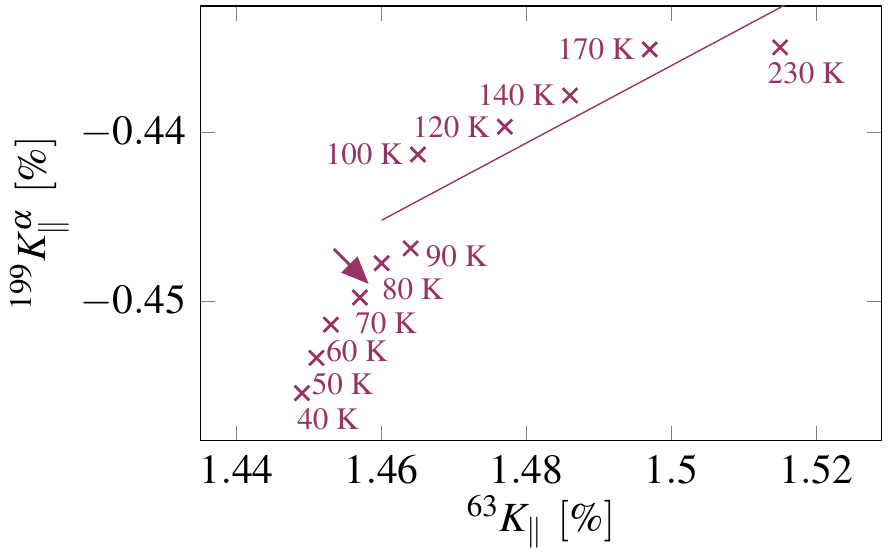}
  \caption{(Color online) $^{199}$Hg shifts $^{199}K_{\parallel}^{\alpha}$ as a function of ${}^{63}$Cu shifts $^{63}K_{\parallel}$ for the UN74 sample with temperature as an implicit parameter, arrow indicates $\tc$ value. Solid line is a fit for $T\geq 90 K$.}
\label{HgCu}
\end{figure}

We begin with the $^{199}$Hg NMR. We plot in Fig.~\ref{HgCu} the shift of the $\alpha$ line (${^{199}K^{\alpha}_\parallel}$) against that of $^{63}$Cu (${^{63}K_\parallel}$) for $\cb$ (for more details see \cite{Haase2012}). For UN74, with no Fermi-liquid-like component present in this sample, we expect a clear break near $T_0 \approx$ 80 K. That is indeed the case. For OP97 the situation is not quite as clear since the Fermi-liquid-like component is not zero. It was proven in Ref. \cite{Haase2012} that even for the Hg resonance  the diamagnetic response in the superconducting state is smaller than half of the total spin shift below about 80 K, so $\chi_\kappa$ must play a role.

\begin{figure}[h]
  \centering
  \includegraphics[width=\columnwidth]{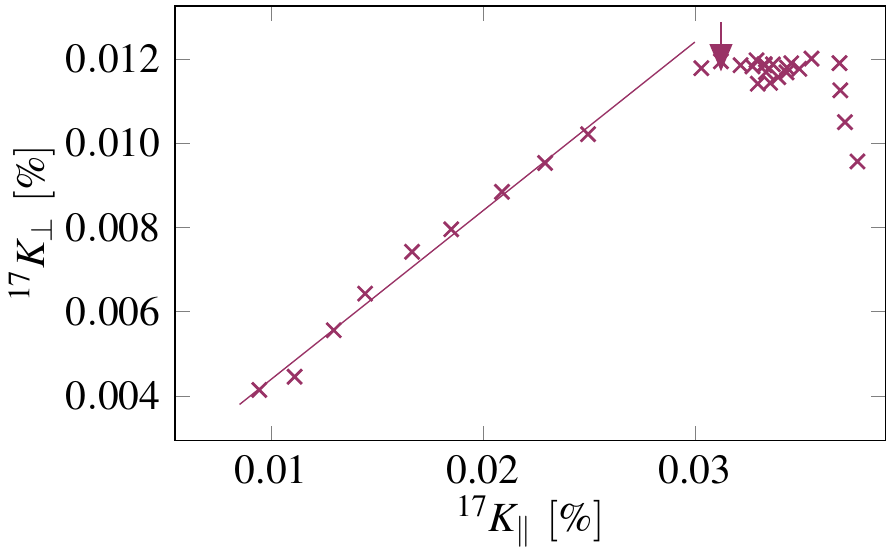}
  \caption{(Color online) Apical oxygen shift $^{17}K_{\bot} $ (O(2), $H_0\parallel a$) as a function of $^{17}K_{\parallel}$ (O(2), $H_0\parallel c$) with temperature as an implicit parameter for a sample with $\tc$ similar to our UN74 sample \cite{Mounce2013} (linear interpolation was applied for $^{17}K_{\bot} (T)$). Solid line is a guide to the eye for shifts below $\tc$ (indicated by an arrow), the slope is 0.4.}
\label{OApexParPerpMounce}
\end{figure}
\begin{figure}[h]
  \centering
  \includegraphics[width=\columnwidth]{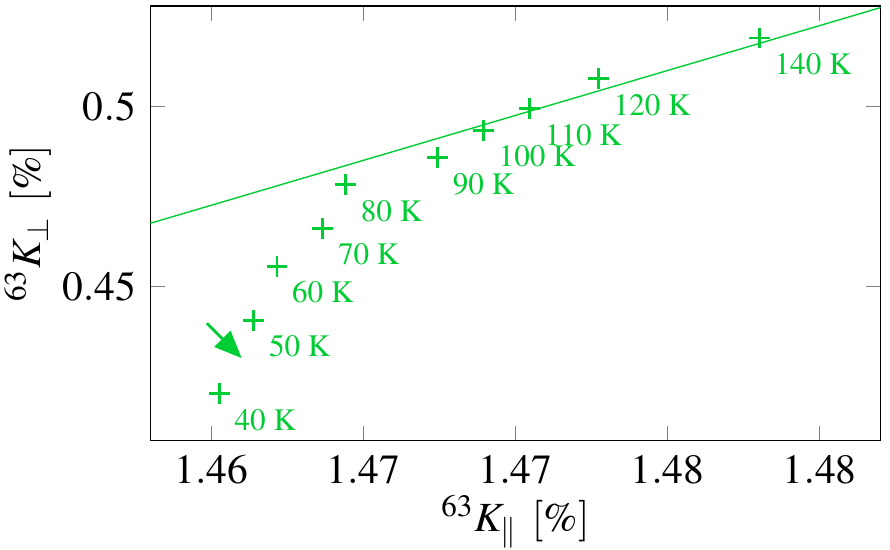}
  \caption{(Color online) ${}^{63}K_{\bot}$ as a function of ${}^{63}K_{\parallel}$ for UN45 sample with temperature as an implicit parameter. A clear deviation from a high temperature linear dependence is observed below 80 K, i.e. well above $\tc$ (indicated by an arrow).}
\label{Cu45}
\end{figure}

An even clearer picture for UN74 emerges if we consider recent $^{17}$O data on the planar and apical oxygen for a sample with very similar doping \cite{Mounce2013}. Apical oxygen data  for $\cb$ and $\cp$ are plotted against each other in Fig.~\ref{OApexParPerpMounce}. We observe a rather small anisotropy below $\tc$ (low temperature slope of $\approx 0.4$). 
This argues against the assumption made by the authors of \cite{Mounce2013} that  the apical oxygen shift decrease below $\tc$ is due diamagnetism only since it leads to an anisotropy ratio of the penetration depth 
$\gamma=\lambda_{\perp}/ \lambda_{\parallel} \approx 6$ while it is expected to be about 40 to 60, an order of magnitude larger \cite{Hofer1998, Panagopoulos2003}. 
Using formulas from \cite{Mounce2013} we can estimate the ratio of the shift decrease below $T_c$, $\Delta K_{\parallel} /  \Delta K_{\perp}$ due to diamagnetism. With demagnetization factors from \cite{Pardo2004} (dimensions of UN74 sample: 1.1 mm x 0.9 mm x 0.5 mm) and for $\gamma$=50 we estimate it to be around 30. However, experimentally for the $^{63}$Cu shifts we obtain the ratio of 0.15.  

The data for UN45 are proof in favor of $\chi_\kappa$ as well since the shift decrease starts far above $\tc$ (Fig.~\ref{Cu45}), ruling out an explanation due to diamagnetism or singlet pairing in the superconducting state.

\begin{figure}[t]
  \centering
  \includegraphics[width=\columnwidth]{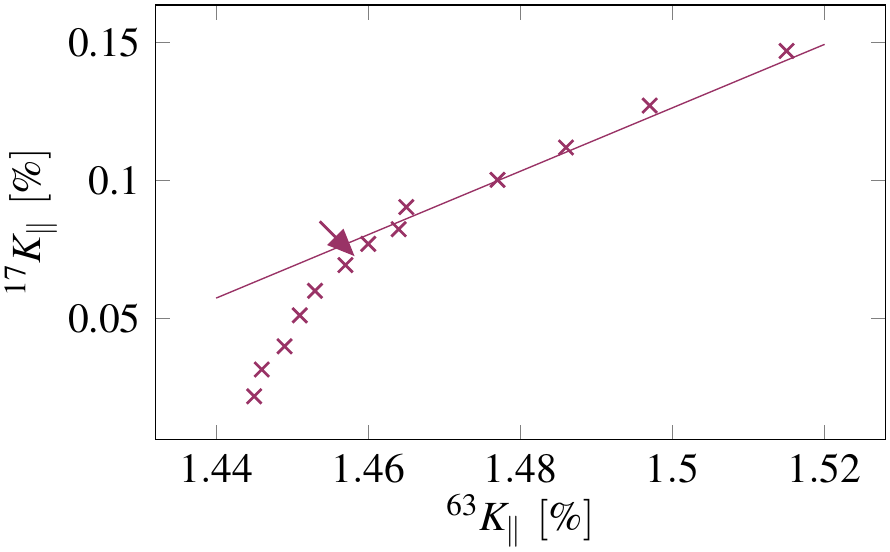}
  \caption{(Color online) Planar oxygen shift $^{17}K_{\parallel}$ (O(1), $H_0\parallel c$) from \cite{Mounce2013} (linear interpolation was applied for $^{17}K_{\parallel}(T)$) as a function of $^{63}K_{\parallel}$ (our results). The reference frequency for ${}^{17}K_{\parallel}$ was chosen such that shift vanishes at $T=0$. Solid line is guide to the eye. Arrow indicates $\tc$.}
\label{OCuMounce}
\end{figure}

In addition, if we look at Fig.~\ref{OCuMounce} where we plot planar oxygen shifts from \cite{Mounce2013} as a function of Cu shifts from our measurements for UN74, the high-temperature slope of about 1.2 gives us the ratio of the hyperfine coefficients of planar O and Cu (for $\cb$) for the pseudogap component, and we find it to be in agreement with the ratio deduced for La$_{1.85}$Sr$_{0.15}$CuO$_4$ (assuming the same $^{63}q_\bot$ for both materials) \cite{Haase2009}. An estimate of the ratio of the hyperfine coefficients for planar O and Cu (for $\cb$) for $\chi_\kappa$ is about -4, very different from what was estimated for the Fermi-liquid-like component (+0.21) \cite{Haase2009}.

To summarize, we believe that there cannot be any doubt that the new shift component is due to a spin susceptibility that is temperature independent at high temperatures, but vanishes at low temperatures, with an onset temperature $T_0$ that can be very different from $\tc$.
\section{Part 2: Numerical Decomposition of Shift Data}
Here we describe the decomposition of the experimentally determined $^{63}$Cu magnetic shifts, $^{63}K_{\parallel,\bot}(T)$, that are displayed in Fig.~1 (main paper), in more detail. The result of the decomposition is shown in Fig.~4 (main paper, for $\cp$). 

If a single electronic spin component  with susceptibility $\chi(T)$ was causing the temperature dependence of the shifts, any nucleus ($n$) would experience a spin shift given by,
\beq
{^{n}K_{S,\parallel\bot}}(T)={^{n}q_{\parallel\perp}}\cdot \chi(T),
\label{eqnComp1}
\eeq
where the anisotropy of the shifts, here just for the two principle axes ($\parallel, \bot$) of the shift tensor (HgBa$_{2}$CuO$_{4+\delta}$ is tetragonal), is given by the anisotropic hyperfine constants (${^{n}q_{\parallel,\bot}}$).
Thus, if we plot the shifts measured at any two nuclei $(n, k)$ against each other (with temperature as an implicit parameter) we must have,
\beq
{^{n}K_{S,\parallel\bot}}(T)=\frac{^{n}q_{\parallel\bot}}{^{k}q_{\parallel\bot}}\cdot {{^{k}K}_{S,\parallel\bot}}(T),
\label{eqnSlope1}
\eeq
and a straight line with the slope given by the ratio of the hyperfine coefficients should emerge. Since this was not the case, e.g., for La$_{1.85}$Sr$_{0.15}$CuO$_4$ with $^{17}K_\parallel$ and $^{63}K_\bot$ \cite{Haase2009}, the shifts were discussed in terms of two susceptibilities $\chi_1$ and $\chi_2$ for two spin components $S_1$ and $S_2$, respectively \cite{Haase2009}:
\beq
{^{n}K_{S,\parallel\bot}}(T)={^{n}q_{_1,\parallel \bot}}\cdot \chi_1(T)+{^{n}q_{2,\parallel \bot}}\cdot \chi_2(T).
\label{eqnComp2}
\eeq
This expression was also used to discuss the $^{17}$O NMR shifts in YBa$_2$Cu$_4$O$_8$ measured as a function of pressure \cite{Meissner2011} and $^{63}$Cu and $^{199}$Hg shifts in two samples of HgBa$_{2}$CuO$_{4+\delta}$ (UN74 and OP97) \cite{Haase2012}. We now know that the shift analyses of the latter samples in terms of \eqref{eqnComp2} are not justified, as there must be another component present in the system ($\chi_\kappa$). Therefore, we have to be  careful when discussing the failure of the single component picture.

If we are only interested in the Cu shift data (from now on we drop the nuclear label), a single-component behavior demands,
\beq
{K_{S,\bot}}(T)=\frac{q_{\bot}}{q_{\parallel}}\cdot{K}_{S,\parallel}(T),
\label{eqnSlope2}
\eeq
and we see with Fig.~2 that Eq.~\eqref{eqnSlope2} does not hold. Consequently we are forced to introduce another susceptibility. However, we cannot just use \eqref{eqnComp2} and we write instead,
\beq
{K_{S,\parallel\bot}}(T)={q_{\parallel\bot}}\cdot \chi(T)+{q_{\kappa,\parallel\bot}}\cdot \chi_\kappa(T),
\label{eqnSysKs}
\eeq
where $\chi$ and $\chi_\kappa$ are two susceptibilities from two different spin components.
Note that Fig.~2 also requires 
\beq
q_\parallel/q_\bot \neq q_{\kappa, \parallel}/q_{\kappa, \bot},
\label{eqnDeter}\eeq
for otherwise Eq.~\eqref{eqnSlope2} would still hold.

For any two temperatures $T_2>T_1$ we write for the corresponding changes in the shift $\Delta K_\eta=K_{\eta}(T_2)-K_{\eta}(T_1)$, and we have with $\Delta \chi=\chi(T_2)-\chi(T_1)$, etc., from \eqref{eqnSysKs},
\begin{align}
  \Delta K_{S\parallel}&=q_{\parallel}\Delta \chi+q_{\kappa\parallel}\Delta \chi_\kappa  \label{eqnSysKs2a}\\ 
  \Delta K_{S\perp}&=q_{\perp}\Delta \chi+q_{\kappa\perp}\Delta \chi_\kappa \label{eqnSysKs2b}.
\end{align}
Note that $\Delta K_\eta \equiv \Delta K_{S\eta}$ since we assumed that the temperature dependence of the total magnetic shift is only due to spin effects. Momentarily, we do not require $K_{S\eta}(T\rightarrow0) \approx 0$, as is typically assumed (for spin singlet pairing).

If we knew the coupling constants $q_\eta$ and $q_{\kappa\eta}$ the system of equations \eqref{eqnSysKs2a}, \eqref{eqnSysKs2b} could be solved for $\Delta\chi$ and $\Delta\chi_\kappa$ (as long as the rank of the corresponding matrix is 2, i.e., the relation \eqref{eqnDeter} is obeyed). However, these values are not accessible from the NMR data alone.

Nonetheless, knowledge of the ratios $q_{\parallel}/q_{\perp}$ and $q_{\kappa\parallel}/q_{\kappa\perp}$ does allow to solve \eqref{eqnSysKs2a} and \eqref{eqnSysKs2b} for the changes of the spin shift $q_\eta\Delta\chi$ and $q_{\kappa\eta}\Delta\chi_\kappa$, i.e.,

\begin{align}
  q_{\parallel}\Delta\chi&=\frac{\Delta K_{\parallel}-\frac{q_{\kappa \parallel}}{q_{\kappa \perp}}{\Delta K_{\perp}}}{1-\frac{q_{\perp}}{q_{\parallel}}\frac{q_{\kappa\parallel}}{q_{\kappa\perp}}}\label{eqnSolutionsDir}\\
  q_{\kappa\parallel}\Delta\chi_\kappa&=\frac{\Delta K_{S\parallel}-\frac{q_{\parallel}}{q_{\perp}}{\Delta K_{\perp}}}{1-\frac{q_{\kappa\perp}}{q_{\kappa\parallel}}\frac{q_{\parallel}}{q_{\perp}}}.\label{eqnSolutionsDirKappa}
\end{align}

Summing up all the incremental changes with temperature we obtain the total temperature-dependent spin shift contributions, i.e., $q_{\eta}\chi_\kappa(T)$ and $q_{\kappa\eta}\chi_\kappa(T)$  ({relative to our starting temperature}).

\begin{figure}[h]
  \centering
  \includegraphics[width=\columnwidth]{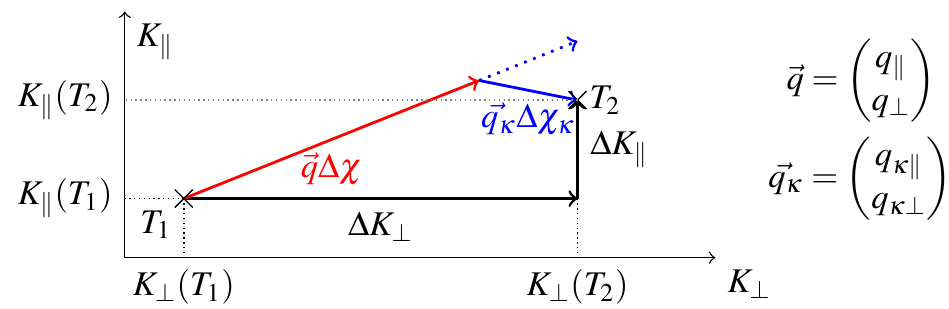}
  \caption{(Color online) Vector interpretation of system of equations \eqref{eqnSysKs2a}, \eqref{eqnSysKs2b}. In the case of two linearly dependent vectors $\vec{q}$ and $\vec{q}_\kappa$, a solution does not exists (dotted line).}
\label{vectorDirections}
\end{figure}

The system of equations \eqref{eqnSysKs2a}, \eqref{eqnSysKs2b} can also be understood in terms of vector addition with relation \eqref{eqnDeter} meaning that the vectors $\left(q_{\parallel}\;\; q_{\perp}\right)$ and $\left(q_{\kappa\parallel}\;\; q_{\kappa\perp}\right)$ are linearly independent, c.f. Fig.~\ref{vectorDirections}.

While $q_{\parallel}/q_{\perp}=0.4$ can be extracted from the common slope at high temperatures in Fig. 2, the ratio $q_{\kappa\parallel}/q_{\kappa\perp}$ cannot be obtained in a similar fashion. Therefore, we use another criterion to find this ratio. When plotting the shifts of UN74 vs. UN45, cf. Fig. 3, we find them to be proportional to each other for $\cb$. For $\cp$  we observe a linear behavior with the same slope above 80~K as for $\cb$. 
This finding hints at a scaling behavior of $K_\eta(x,T)$ due to that of $\chi(x,T)$ and that the deviation from it at lower temperatures is caused by $q_{\kappa\eta}\chi_\kappa(T)$. We postulate that this scaling  of $\chi(x,T)$ is present in the whole temperature range of our measurements. It turns out that this scaling is even in quantitative agreement with an extensive body of susceptibility data by Johnston \cite{Johnston1989} and Nakano et al.~\cite{Nakano1994} as will be shown in Part 3.

\begin{figure}[h]
  \centering
  \includegraphics[width=\columnwidth]{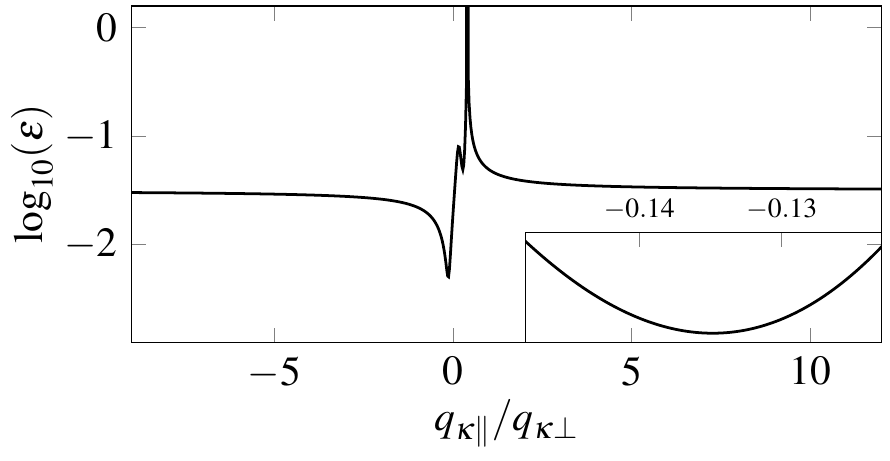}
  \caption{(Color online) Logarithmic error $\epsilon$ of linear fit according to Eq.~\eqref{eqScalingChi} as a function of $q_{\kappa\parallel}/q_{\kappa\perp}$.}
\label{qTest}
\end{figure}

We perform the decomposition according to Eqs.~\eqref{eqnSolutionsDir} and \eqref{eqnSolutionsDirKappa} with $q_{\kappa\parallel}/q_{\kappa\perp}$ as one parameter for both underdoped samples. For each value of this parameter, we then apply a linear regression of $q_{\eta}\chi^\text{UN74}$ vs. $q_{\eta}\chi^\text{UN45}$ and calculate the error $\epsilon$ as the sum the squares of all temperature points,
\begin{align}
  q_{\eta}\chi^\text{UN74}(T)= \underbrace{\frac {x^\text{UN74}} {x^\text{UN45}}}_{=a} \cdot q_{\eta}\chi^\text{UN45}(T)\label{eqScalingChi}\\
  \epsilon=\sqrt{\sum_{T_i}\left(q_{\eta}\chi^\text{UN74}(T_i)-a\cdot q_{\eta}\chi^\text{UN45}(T_i)\right)^2},
\end{align}
with $a=1.7 \pm 0.2$ \cite{Barisic2008} (we allow the doping ratio to change within the error). 
The value that gives the best agreement in Fig.~\ref{qTest} is $q_{\kappa\parallel}/q_{\kappa\perp}=-0.135$. We use this value for the decomposition of the shift data of all samples.

\begin{figure}[h]
  \centering
  \includegraphics[width=\columnwidth]{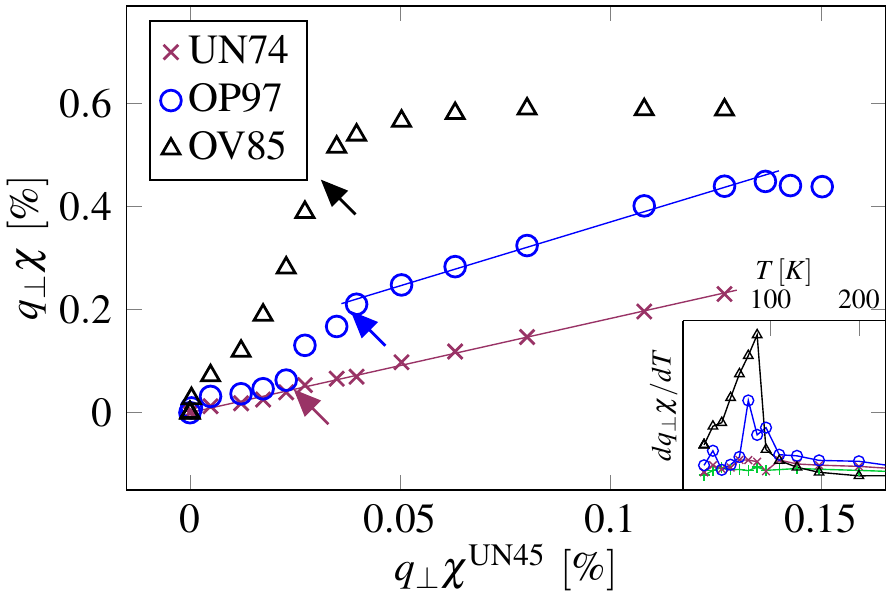}
  \caption{(Color online) Scaling of the $\chi$ susceptibility obtained from NMR measurements. $q_{\bot} \chi$ of the UN74, OP97 and OV85 samples plotted versus $q_{\bot} \chi$ of the UN45 sample with temperature as an implicit parameter. Solid lines are fit to the data points with slopes that agree with the doping ratios within their margin of error. Inset shows derivative $d(q_{\bot} \chi) / dT$ as a function of temperature.}
\label{scalingChi}
\end{figure}

The shift component ($q_{\eta}\chi$) that remains after the subtraction of $q_{\kappa \eta} \chi_\kappa$ differs widely between the samples. For the two underdoped samples (UN45 and UN74) we find $q_{\eta}\chi(T)$  to change slowly with temperature even throughout $\tc$, i.e., $\tc$ is not noticeable, cf. also inset of Fig.~\ref{scalingChi}. This is very different for OP97 and in particular for OV85 where this shift component changes rapidly below $\tc$. The question arises whether one can explain such a behavior with only one temperature-dependent spin component. 
It was shown by comparing the shifts of different nuclei for this system \cite{Haase2012}, as well as for other systems \cite{Haase2009, Meissner2011}, that a Fermi-liquid-like component is present, i.e., a shift component that is temperature-independent above $\tc$ and vanishes rapidly below it. Fermi-liquid-like behavior is seen in the bare shift data, cf. Fig. 1 (main text) for the optimal and overdoped samples, as well as in Fig.~\ref{scalingChi}. However, it is not visible when we plot parallel and perpendicular shifts against each other, cf. Fig.~2 (main paper). This means that if both spin components are present they must couple to the Cu nucleus with the same anisotropy ratio of hyperfine coefficients (most likely the same coefficients, as if the spins couple through the same atomic orbital). We will assume that a Fermi-liquid-like component ($\chi_2(T)$) is present. We will see below that this defines a $\chi_1 (T)$ that obeys the scaling property found for many systems early on. Therefore, we write
\begin{align}
  q_{\eta}\chi(T)&=q_{\eta}\left[\chi_1(T)+\chi_2(T)\right].
  \label{flandpg}
\end{align}

For UN45 and UN74 the Fermi-liquid-like component must vanish ($\chi_2(T)\approx 0$) as noticed in Fig.~\ref{scalingChi}. We also notice in this figure the scaling behavior that was mentioned in the main text, and that it applies to $\chi_1(T)$, only. We will assume that this scaling behavior of $\chi_1(T)$ also holds for the other two samples up to a certain, sample-dependent temperature (which follows from the scaling below, about half the characteristic temperature of pseudogap component $T_{max}$, see Part 3 in this Supplement). This means, we determine $\chi_1(T)$ for UN74, OP97, and OV85 from  $\chi_1^\text{UN45}(T) (=\chi^\text{UN45}(T))$, according to $\chi_1^j(T)=x^j/x^\text{UN45}\cdot \chi^\text{UN45}(T)$, where $x^j$ denotes the doping level.
With $\chi_1(T)$ we also have $\chi_2(T)=\chi(T) - \chi_1 (T)$. Since we assume that the scaling will break down for high doping levels at high temperatures where the Fermi-liquid-like component dominates, we define that $\chi_2(T\gtrsim1.1 \cdot \tc)=const.$ This means we force the temperature-dependent shift above $\tc$ to be due to $\chi_1(T)$ (we will see in Part 3 that this yields a $\chi_1(T)$ in agreement with the pseudogap susceptibility). The analysis gives $\chi_1(T)$ and $\chi_2(T)$ as shown in Fig.~4 (main paper).
\vspace{0.5cm}
\section{Part 3: Scaling of $\chi_1$}
We now focus on the extracted pseudogap shift given by $q_{1,\eta} \chi_1(T)$. Based on shift and linewidth data for UN74 and OP97 we concluded previously \cite{Rybicki2009, Haase2012} on a scaling property for the pseudogap susceptibility, $\chi_1(x^j,T)=x^j/x^i \chi_1(x^i,T)$. The new results for UN45 confirm this scaling. 

An experimental scaling relationship for part of the uniform spin susceptibility was introduced by Johnston \cite{Johnston1989} based on measurements on powder samples of La$_{2-x}$Sr$_x$CuO$_4$ and were confirmed with large sets of measurements on this system \cite{Nakano1994} and on Y-doped Bi$_2$(Sr,Ca)$_3$Cu$_2$O$_8$ \cite{Oda1990}. Uniform susceptibility measurements can be reliably performed only above $\tc$, and the authors showed that all their data (also on different materials) can be fit with a sum of two susceptibilities: a pseudogap susceptibility with a universal temperature dependence and a temperature independent, but doping dependent susceptibility that was assumed to be a Fermi liquid contribution. Since it is difficult to discern a Fermi liquid term (above $\tc$) from orbital effects, the Fermi liquid part was assumed to be zero for zero doping. We will show here that the temperature-dependent pseudogap susceptibility described by the universal function is in quantitative agreement with our proposed $\chi_1(x,T)$ for very similar characteristic temperatures. 

\begin{figure}[h]
  \centering
  \includegraphics[width=\columnwidth]{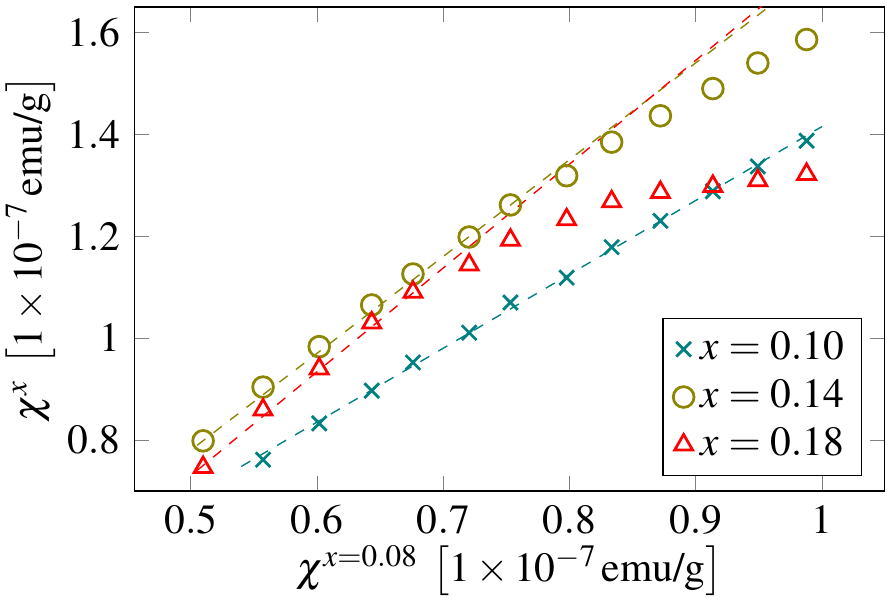}
  \caption{(Color online) Scaling of $\chi$ in \lsco{} deduced from uniform susceptibility measurements \cite{Nakano1994}. Data are shown in temperature range similar to ours (50 K - 300 K). Dashed lines are a guide to the eye.}
\label{figRawNakanoSuscScaling}
\end{figure}

In Fig.~\ref{figRawNakanoSuscScaling} we show data from \cite{Nakano1994}: we plot the susceptibilities of the samples with higher doping against that of a sample with $x$=0.08 (compare to our shifts shown in Fig.~3). Linear dependencies are observed for the underdoped samples, as well for low enough temperatures at higher doping levels, and the slopes agree with the doping ratios. Such a similarity in very different cuprates is remarkable and must mean that the pseudogap susceptibility is indeed ubiquitous to the cuprates.

\begin{figure}[h]
  \centering
  \includegraphics[width=\columnwidth]{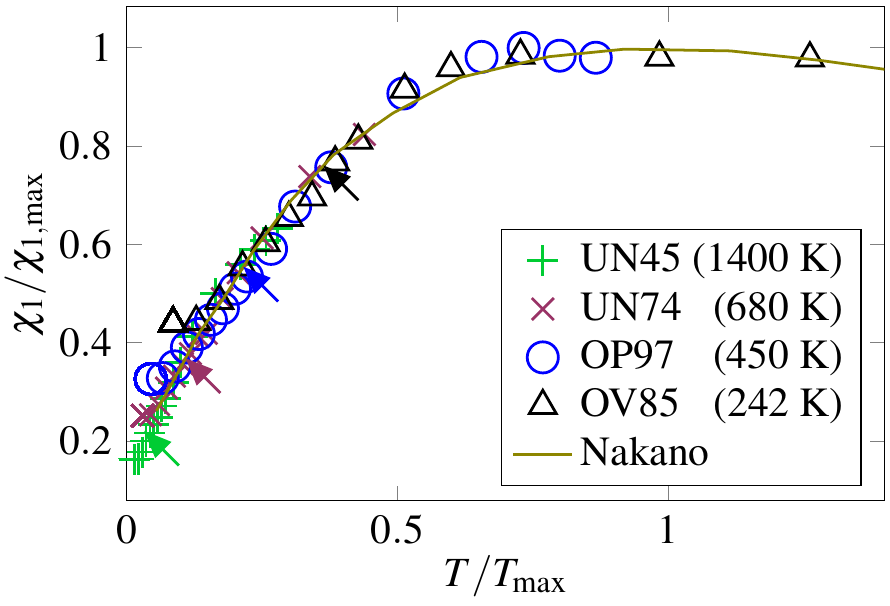}
  \caption{(Color online) Normalized pseudogap susceptibility i.e. $\chi_1(T) / \chi_{1,max}$ versus $T/T_\text{max}$ for all studied samples of HgBa$_{2}$CuO$_{4+\delta}$ with the universal curve (solid red line) derived by Nakano et al. \cite{Nakano1994}. $T_\text{max}$ values are given in parenthesis, arrows indicate $\tc$ for given sample.}
\label{nakano}
\end{figure}

The pseudogap susceptibilities all fit the universal function $F(T/T_\text{max}(x))$ shown in Fig.~\ref{nakano}, which has two parameters, the doping-dependent characteristic temperature $T_\text{max}(x)$ and $\chi_\text{max}(x)$, the maximum of the pseudogap susceptibility at $T=T_{max}$. {This universal curve shows an almost linear behavior at low temperatures that is our observed scaling behavior.} The scaling breaks down at temperatures $\approx T_\text{max}/2$ (both parameters have been tabulated for the various doping levels in \cite{Nakano1994}).

Since we do not have measurements at very high temperatures, we use values of $T_\text{max}$ for UN45, UN74 and OP97 for the same doping as those deduced by Johnston and Nakano et al. for La$_{2-x}$Sr$_x$CuO$_4$. For the overdoped OV85 sample we can calculate $T_\text{max}$ reliably from our data and we obtain 242(4) K. 
We note that our $q_{1,\eta}\chi_1(x,T)$ is in almost perfect agreement with the universal curve $F(T/T_\text{max}(x))$ as shown in Fig.~\ref{nakano}. 

We stress that $F(T/T_\text{max}=0)\neq 0$. So it might be that there is a pseudogap shift even at zero temperature. 
We cannot know whether this is true or just a result of the assumptions for the temperature independent part.
It is also important to point out that the temperature independent susceptibility above $\tc$ that was ascribed to a Fermi liquid component by Johnston and Nakano must contain our $\chi_\kappa$. Inspection of the susceptibility data shows that this is possible. We also note that low temperature susceptibility data are less reliable due to a seemingly large contributions from free Curie spins \cite{Johnston1989,Nakano1994}.

We conclude that the scaling of the NMR shifts that we also reported for YBa$_2$Cu$_4$O$_8$ \cite{Meissner2011} must be generic to the cuprates and reflects the properties of the pseudogap susceptibility. It is also clear that a second, Fermi liquid-like component is present in all cuprates. However, we cannot say from the data at hand if $\chi_\kappa$ is ubiquitous to the cuprates, as well. Nevertheless, given the overall shift behavior of HgBa$_{2}$CuO$_{4+\delta}$ that is representative for most of the cuprates, we believe that $\chi_\kappa$ must be present in other materials as well.
\bibliography{HTSCsupp}
\bibliographystyle{apsrev4-1}